\documentclass[journal,10pt]{IEEEtran} 
\usepackage{amssymb}
\usepackage{amsmath}
\usepackage{bm}
\usepackage{amsfonts}
\usepackage{graphicx}
\usepackage{subfigure}
\usepackage{cite}
\usepackage{enumerate}
\usepackage{url}
\usepackage{setspace}
\usepackage[ruled,linesnumbered]{algorithm2e}
\usepackage{makecell}
\usepackage{mathtools}
\usepackage{epstopdf}
\usepackage{dsfont}
\usepackage[usenames,dvipsnames]{color}
\usepackage{colortbl}
\definecolor{tianlan}{rgb}{.94, 1, 1}
\definecolor{qianhuilan}{rgb}{.69,.88,.90}
\definecolor{xiangyabai}{rgb}{.98, 1,.94}
\definecolor{baixinren}{rgb}{1, .92, .80}
\definecolor{danke}{rgb}{.99, .90, .79}
\definecolor{milucheng}{rgb}{.94, 1, .94}
\definecolor{huang1}{rgb}{1, .98, .84}
\definecolor{lan1}{rgb}{.88, .92, .98}
\usepackage{makecell}
\usepackage{multirow}
\usepackage[table,xcdraw]{xcolor}
\usepackage{makecell}

\allowdisplaybreaks[4]
\begin{document}

\title{Reconfigurable Holographic Surface: A New Paradigm to Implement Holographic Radio}
\author{
\IEEEauthorblockN{
\normalsize{Ruoqi Deng,~\IEEEmembership{Student Member,~IEEE}},
\normalsize{Yutong Zhang,~\IEEEmembership{Student Member,~IEEE}},
\normalsize{Haobo Zhang,~\IEEEmembership{Student Member,~IEEE}},\\
\normalsize{Boya Di,~\IEEEmembership{Member,~IEEE}},
\normalsize{Hongliang Zhang,~\IEEEmembership{Member,~IEEE}},
\normalsize{and Lingyang Song,~\IEEEmembership{Fellow,~IEEE}}\\}
\vspace{-0.8cm}

\thanks{Ruoqi Deng, Yutong Zhang, Haobo Zhang, Boya Di, and Lingyang Song are with School of Electronics, Peking University, Beijing, China. Lingyang Song is also with Peng Cheng Laboratory, China (email: ruoqi.deng@pku.edu.cn; yutongzhang@pku.edu.cn; haobo.zhang@pku.edu.cn; boya.di@pku.edu.cn; lingyang.song@pku.edu.cn).}
\thanks{Hongliang Zhang is with Department of Electrical and Computer Engineering, Princeton University, NJ, USA (email: hz16@princeton.edu).}
}
\maketitle

\begin{abstract}
Ultra-massive  multiple-input multiple-output (MIMO) is one of the key enablers in the forthcoming 6G networks to provide high-speed data services by exploiting spatial diversity. In this article, we consider a new paradigm termed holographic radio for ultra-massive MIMO, where numerous tiny and inexpensive antenna elements are integrated to realize high directive gain with low hardware cost. We propose a practical way to enable holographic radio by a novel metasurface-based antenna, i.e., reconfigurable holographic surface (RHS).  Specifically, RHSs incorporating densely packed tunable metamaterial elements are capable of holographic beamforming. Based on the working principle and hardware design of RHSs, we conduct full-wave analyses of RHSs and build an RHS-aided point-to-point communication platform supporting real-time data transmission. Both simulated and experimental results show that the RHS has great potential to achieve high directive gain with a limited size, thereby substantiating the feasibility of RHS-enabled holographic radio. Moreover, future research directions for RHS-enabled holographic radio are also discussed.

\end{abstract}



\vspace{-0.4cm}
\section{Introduction}
The sixth generation (6G) wireless communications look forward to providing high-speed data services and revolutionary mobile connectivity to handle the explosive growth in the number of mobile devices and applications. By exploiting the spatial diversity, massive multiple-input multiple-output (MIMO) with large-scale phased arrays capable of highly directional beamforming is once considered one of the powerful solutions to fulfill the challenging visions of 6G communications~\cite{EOF-2014}. However, considering that phased arrays rely on high-resolution phase shifters, the hardware cost and power consumption will become unaffordable as massive MIMO evolves into ultra-massive MIMO~\cite{ZZF-2021}.


To overcome the limitations of phased arrays, a new paradigm termed \emph{holographic radio} for ultra-massive MIMO has been proposed, where high directive gain can be achieved by numerous integrated antenna elements with low hardware cost~\cite{ATL-2020}. To lift the half-wavelength restriction and enable holographic radio in practical systems, \emph{reconfigurable holographic surfaces (RHSs)}, one of the representative metasurface-based antennas composed of densely packing sub-wavelength metamaterial elements are developed as a promising candidate~\cite{BFY-2020}. Specifically, due to the unique structure and characteristics of the metamaterial elements, the RHS can regulate electromagnetic waves via simple diode-based controllers. This provides a powerful solution to reduce the hardware cost while guaranteeing high directive gain in practice.

Different from another widely-used metasurface-based antenna named reconfigurable intelligent surface (RIS) whose feeds are set outside the meta-surface due to the reflection characteristic~\cite{BHL-2020}, the feeds of the RHS are attached to the edge of the metasurface. As such, electromagnetic waves generated by the feeds propagate along the RHS elements and excite the RHS elements one by one. This enables the RHS to serve as an antenna array integrated with the transceiver conveniently. Compared with parallel feeding adopted by phased arrays where each antenna element requires a long feeding line, such series feeding adopted by the RHS also leads to a much simpler wire layout in the implementation of ultra-massive MIMO.  Based on the holographic interference principle, each RHS element can control the radiation amplitude of the incident electromagnetic waves electrically to generate object beams~\cite{MSJ-2014}, and such a beamforming technology is also known as \emph{holographic beamforming}.

Although RHSs provide a promising solution to enable the holographic radio, as nascent meta-surfaces, initial research on RHSs has primarily focused on the fundamental hardware component design~\cite{T-2016} and holographic beamforming optimization~\cite{RBS-2021},~\cite{NOY-2019}. However, most works have only substantiated the capability of a simple one-dimensional RHS to achieve beamforming or investigated RHS-aided communications as well as amplitude-controlled beamforming at the theoretical level. The realization of the RHS-enabled holographic radio system has not been studied. In this article, we study the feasibility of utilizing RHSs to enable holographic radio at the system level and component level. More precisely, we contribute to the research on RHS-enabled holographic radio from the following two aspects:
\begin{itemize}
\item \emph{Verify the potential of RHSs to implement ultra-massive MIMO:} We introduce amplitude-controlled holographic beamforming and present the hardware design of RHSs. Full-wave analyses of both single-feed one-dimensional RHS and multi-feed two-dimensional RHS are then conducted. Simulation results prove that benefitting from the diode-based controller and the series feeding method, the RHS has great potential to achieve high directive gain with low hardware cost and a simple wiring layout, thereby providing a practical way toward ultra-massive MIMO.
\item \emph{Realize a prototype of the RHS-enabled holographic radio system:} To substantiate the feasibility of RHS-enabled holographic radio, we implement an RHS prototype and measure both transmit and receive beam patterns. The experimental results show good agreement with the simulation results and validate the transceiver reciprocity of the RHS. An RHS-aided point-to-point communication platform is then built and is proved to support real-time data transmission.
\end{itemize}

Although RHS-enabled holographic radio provides a powerful solution for the implementation of ultra-massive MIMO, several open problems still need to be addressed, which shed light on future research directions as listed below.
\begin{itemize}
\item \emph{Satellite networks:} RHSs can be integrated with user terminals (UTs) conveniently to support satellite communications due to their ultra-thin and lightweight structures. Since existing algorithms optimizing traditional complex-valued analog beamformer do not work well for real-valued holographic beamformer, a new holographic beamforming optimization is thus required. In addition, an RHS-aided multi-satellite communication protocol also needs to be designed to cope with the high mobility of satellites.
\item \emph{Integrated sensing and communication (ISAC):} ISAC can mitigate the spectrum congestion issue by making sensing and communication systems share the common spectrum. Since the RHS has the potential to achieve high gain with a limited size, RHS-enabled holographic radio can be applied in ISAC systems. Hence, a holographic beamforming algorithm jointly considering the sensing and communication functionalities is required.
\emph{Wireless simultaneous localization and mapping (SLAM):} Wireless SLAM using antennas to exploit the multipath effect of signals provides a promising solution for location and sensing-based services. Since the RHS has a superior beam-steering capability with numerous integrated elements, RHS-enabled holographic radio can be utilized in wireless SLAM systems to improve positioning accuracy.
\end{itemize}

The rest of this article is organized as follows. In Section \uppercase\expandafter{\romannumeral2}, the working principle of RHSs is introduced. In Section \uppercase\expandafter{\romannumeral3}, the hardware design and full-wave analyses of RHSs are presented. The experimental prototype of RHS-enabled holographic radio and experimental results are illustrated in Sections \uppercase\expandafter{\romannumeral4} and \uppercase\expandafter{\romannumeral5}, respectively. Future research directions for RHS-enabled holographic radio are discussed in Section \uppercase\expandafter{\romannumeral6}. Finally, we draw the conclusions in Section \uppercase\expandafter{\romannumeral7}.

\section{Working Principle of Reconfigurable Holographic Surfaces}

In this section, we present the holographic principle of RHSs, based on which amplitude-controlled holographic beamforming is introduced.

\subsection{Holographic Principle}
An RHS is a special leaky-wave antenna comprising feeds and metamaterial radiation elements. As shown in Fig. 1, The feeds are attached to the edge or the back of the RHS and transform input signals into electromagnetic waves, which are also called reference waves. The RHS adopts series feeding where the incident reference wave propagates along the RHS elements and excites the RHS elements one by one. The reference wave is then transformed into a leaky wave through the slots of RHS elements to emit signals to free space~\cite{DO-2016}. The radiation amplitude of the leaky wave at each RHS element can be controlled electrically to achieve holographic beamforming. Compared with parallel feeding adopted by phased arrays where each antenna element requires a long feeding line, the series feeding leads to a much simpler wiring layout in the implementation of ultra-massive MIMO.

To further interpret the working principle of an RHS, we assume that the RHS has $K$ feeds and $N$ RHS elements. The object wave propagating in the direction $(\theta, \phi)$ is denoted as $\Psi_{obj}$, while the reference wave generated by feed $k$ is denoted as $\Psi_{ref}$. The RHS can construct a \emph{holographic pattern} on the metasurface to record the interferogram $\Psi_{intf}$ between the reference wave and the object wave based on the holographic principle, i.e., $\Psi_{intf}=\Psi_{obj}\Psi_{ref}^{\ast}$~\cite{MSN-2016}. When the interferogram $\Psi_{intf}$ recorded on the RHS is excited by the propagating reference wave, the leaked wave $\Psi_{intf}\Psi_{ref}$ is proportional to
$\Psi_{obj}|\Psi_{ref}|^2$, indicating that its direction is exactly $(\theta, \phi)$.

\begin{figure}[t]
\centering
\includegraphics[width=2.9in]{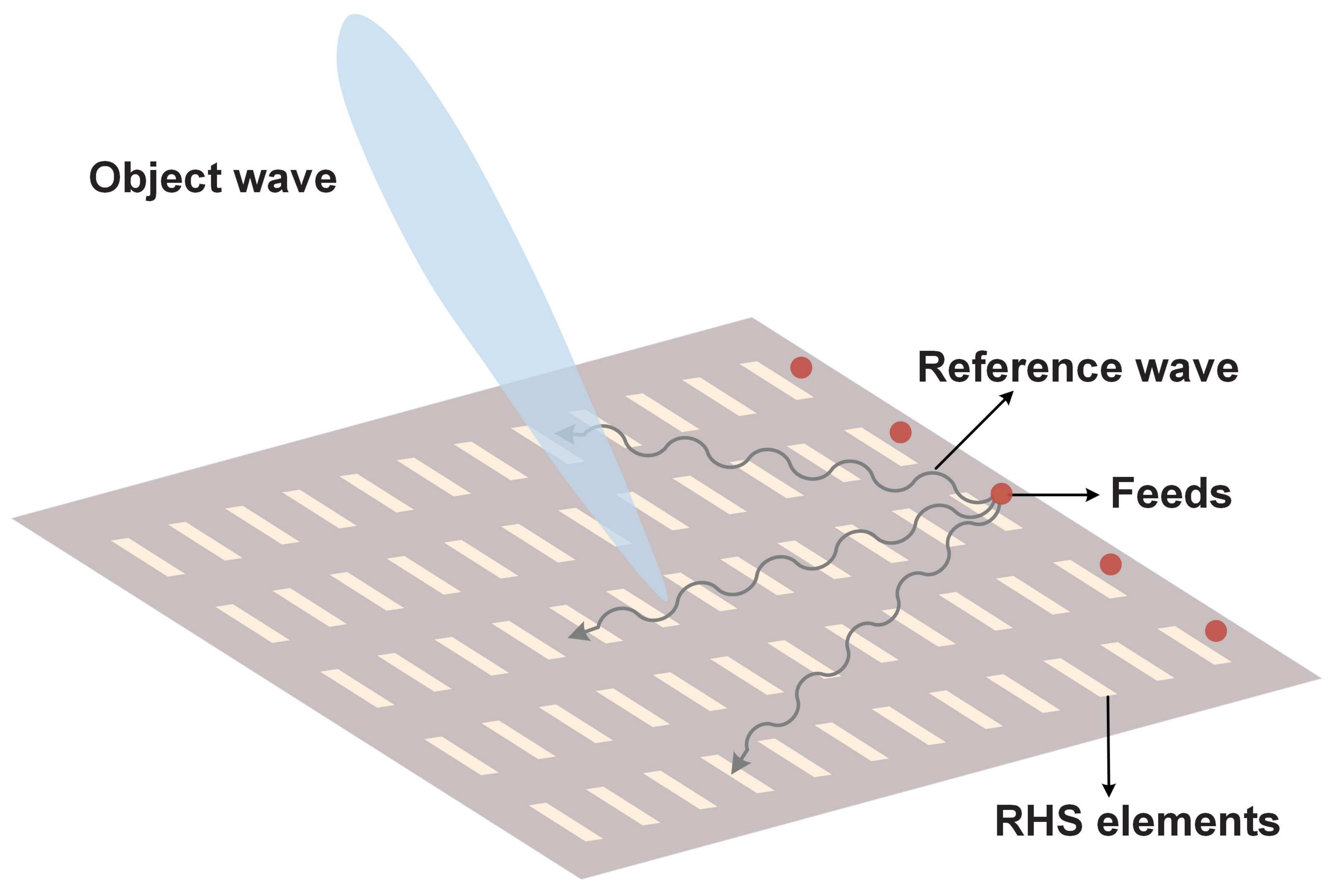}
\caption{Illustration of the RHS.}
\vspace{-0.4cm}
\label{fig1}
\end{figure}

\textbf{Remark 1.} \emph{The physical structure and operating mechanism of RHS are different from those of another representative metasurface called RIS. Specifically, an RHS's feeds are attached to the edge of the metasurface. Hence, RHSs can be integrated with transceivers conveniently. In contrast, the feeds of an RIS are set outside the metasurface owning to the reflection characteristic~\cite{BHL-2020}. An extra link is required between the transmitter and the RIS to control the phase shifts of each RIS element. In addition, the RHS utilizes the method of series feeding, while the RIS utilizes the method of parallel feeding where all RIS elements are excited by the incident signals at the same time. Due to their different physical structure and operating mechanism, RHSs are more likely to serve as transmit/receive antennas directly, while RISs are widely used as relays.}


\subsection{Amplitude-Controlled Holographic Beamforming}

The RHS utilizes an amplitude-controlled method to represent the information contained in the interferogram $\Psi_{intf}$ by tuning each RHS element based on the phase of the reference wave at each RHS element. Specifically, the phase of the reference wave changes in the process of propagation. At each RHS element, the phase of the reference wave is determined by the product of the propagation vector on the RHS and the distance vector from feed $k$ to the $n$-th RHS element $\mathbf{r}_n^k$~\cite{MSN-2016}, which is a-priori fixed. If the reference wave is in phase with the object wave at an RHS element, the RHS element will be tuned to radiate much energy of the reference wave into free space. Otherwise, the RHS element will be detuned and not radiate energy into free space.

To map the phase difference between the object wave and the reference wave to the radiation amplitude of the RHS elements, the real part of the interference $\text{Re}[\Psi_{intf}]$ (i.e., the cosine value of the phase difference) is considered. Since the value of $\text{Re}[\Psi_{intf}]$ decreases as the phase difference grows, satisfying the amplitude control requirements, $\text{Re}[\Psi_{intf}]$ is a direct representation of the radiation amplitude distribution on the RHS. Hence, the holographic pattern $\bm{m}$, i.e., the radiation amplitude of each RHS element which can generate the object wave with the direction of $(\theta, \phi)$ is parameterized mathematically by
\begin{equation}\label{M}
m(\mathbf{r}_n^k, \theta, \phi)=\frac{\text{Re}[\Psi_{intf}(\mathbf{r}_n^k, \theta, \phi)]+1}{2},
\end{equation}
where $\text{Re}[\Psi_{intf}]$ is normalized to [0,1] to avoid negative values. This formula represents the basic principle for amplitude-controlled holographic beamforming.  The effectiveness of this formula to achieve holographic beamforming will also be verified in Section~\uppercase\expandafter{\romannumeral5}.


\vspace{-0.1cm}
\section{Hardware Design and Full-wave analyses of Reconfigurable Holographic Surfaces}
In this section, we present the hardware design of the RHS element. Full-wave analyses of RHSs are then introduced.

\begin{figure}[t]
\centering
\includegraphics[width=3in]{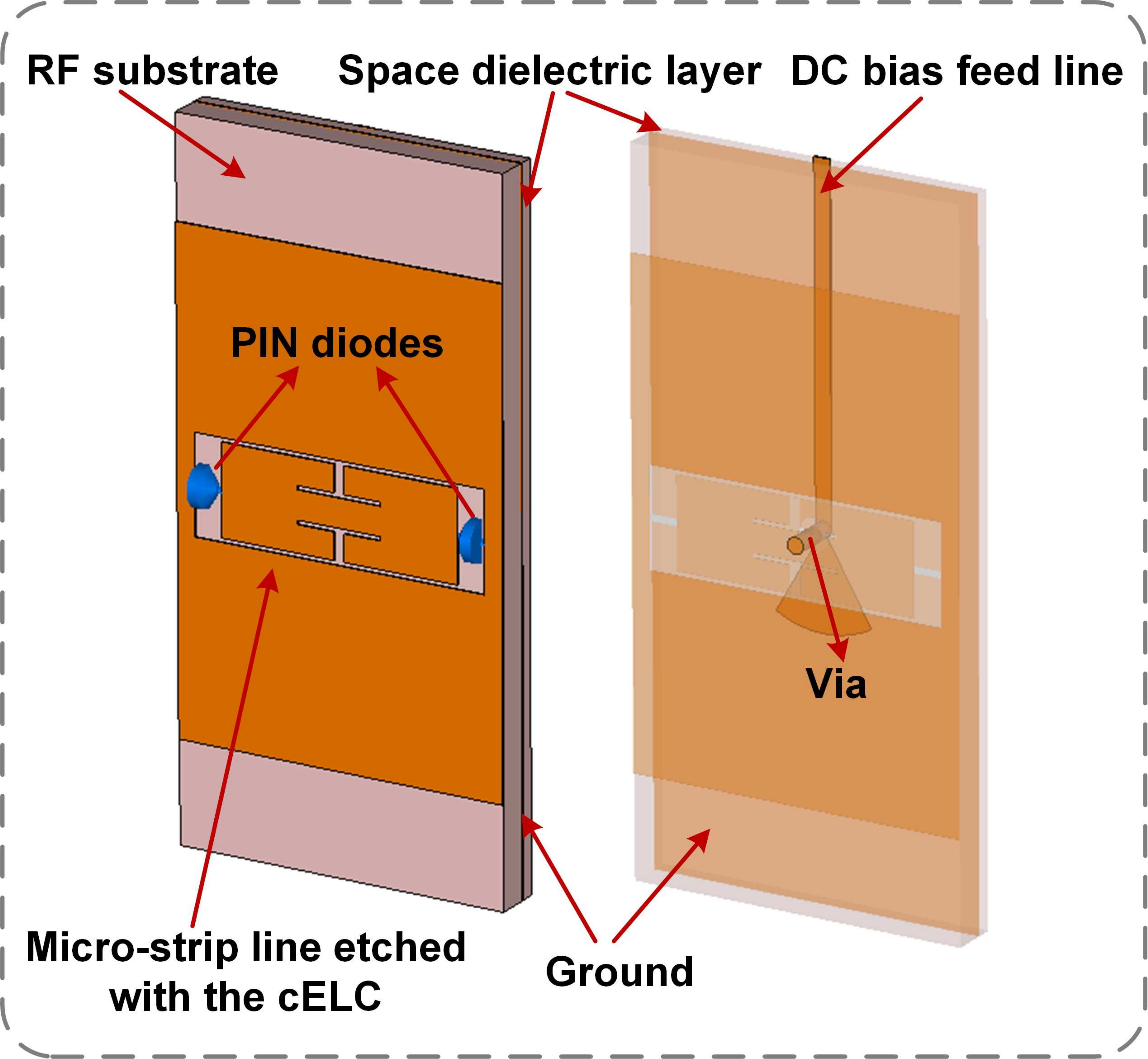}
\caption{Hardware design of the RHS element.}
\vspace{-0.4cm}
\label{fig1}
\end{figure}

\vspace{-0.2cm}
\subsection{Hardware Design}
As shown in Fig. 2, the kernel of the designed RHS element with controllable radiation amplitude is a complementary electric-LC (cELC) resonator connected with PIN diodes\footnote{The RHS element with controllable radiation amplitudes can also be fabricated by loading varactor diodes and liquid crystals~[6].}~\cite{T-2016}. For symmetry, two PIN diodes are connected across the gaps separating the central metal patch from the micro-strip line. By controlling the biased voltages applied to the PIN diodes, the cELC resonator's mutual inductance together with the radiation amplitude of the RHS element can be changed. Specifically, when two PIN diodes are in the OFF states, the RHS element radiates the energy of the reference wave into free space. When the PIN diodes are in ON states, the reference wave's energy is hardly radiated\footnote{The design requirement of the radiation efficiency of an RHS element is related to the size of the RHS to guarantee that  most of the input energy from the feed can be radiated into free space. For example, for a one-dimensional RHS with 16 RHS elements, the radiation efficiency of an RHS element with PIN diodes in the OFF and ON states is required to be 30\%-40\% and lower than 15\%, respectively~\cite{T-2016}.}.

The size of the RHS element is $0.82\times 1.7\times 0.11$ $\text{cm}^3$. The aimed working frequency is set as 12~GHz for satellite communications. To cover the working frequency, we choose MACOM MADP-000907-14020 with low insertion loss for PIN diodes~\cite{HY-2016}. The whole RHS element is composed of five layers given as below.
\begin{itemize}
 \item \emph{The top layer} is the micro-strip line etched with the cELC.
 \item \emph{The second layer} is the radio frequency (RF) substrate, which is 0.8-mm-thick F4B, for the propagation of the reference wave.
 \item \emph{The third layer} is the ground plane.
 \item\emph{The fourth layer} is the spacer dielectric layer, which is 0.2-mm-thick F4B.
 \item\emph{The bottom layer} is the direct-current (DC) bias feed line which can apply biased voltages to the PIN diodes through a via extending through the cELC, the RF substrate, the ground plane, and the spacer dielectric layer.
\end{itemize}

Moreover, the radiation characteristics of an RHS element such as its radiation efficiency and resonant frequency are determined by its geometric parameters. Considering that the geometric parameters of the RHS element are coupled with each other and cannot be adjusted independently through a single dimension, a parameter optimization procedure based on full-wave analyses is also applied to achieve the desired radiation characteristics.


\vspace{-0.2cm}
\subsection{Full-Wave Analyses of RHSs}
Based on the simulation of the RHS element, we conduct full-wave analyses of RHSs with different sizes utilizing the CST Microwave Studio.

\subsubsection{One-Dimensional RHS}
To begin with, we consider a one-dimensional RHS with 16 RHS elements\footnote{The experimental prototype of the considered one-dimensional RHS will be presented in Section~\ref{EP}.}. The element spacing of the RHS is $0.82$ cm, which is approximately one-third of the wavelength at 12 GHz. The RHS is fed from the port at the left end, where the electromagnetic wave generated by port~1 continuously radiates energy from each RHS element in the propagation process. The residual energy of the electromagnetic wave will be absorbed by the port at the right end.

To determine the ON/OFF state of the PIN diodes of each RHS element, for a given object beam, we calculate the theoretical radiation amplitude of each RHS element by~(1). If the radiation amplitude is larger than a predefined threshold, the PIN diodes will be in the OFF state. In contrast, if the radiation amplitude is less than the threshold, the PIN diodes will be in the ON state~\cite{RD-2021}. The full-wave analysis shows the main lobe's direction of the beam pattern of the RHS matches the object beam's direction and the simulated gain is about 4~dBi\footnote{The gain of the RHS will be slightly different with the change in the main lobe's direction.}, which is obtained directly from the far-field results in the CST Microwave Studio.

\subsubsection{Two-Dimensional RHS}
Considering that a two-dimensional RHS can achieve three-dimensional holographic beamforming, i.e., the elevation angle $\theta$ and the azimuth angle $\phi$ of the generated beam can be controlled, we evaluate the beam-steering capability of the RHS in both horizontal plane and vertical plane.


\begin{figure}[t]
\centering
\subfigure[Horizontal plane]{
	\includegraphics[width=2.7in]{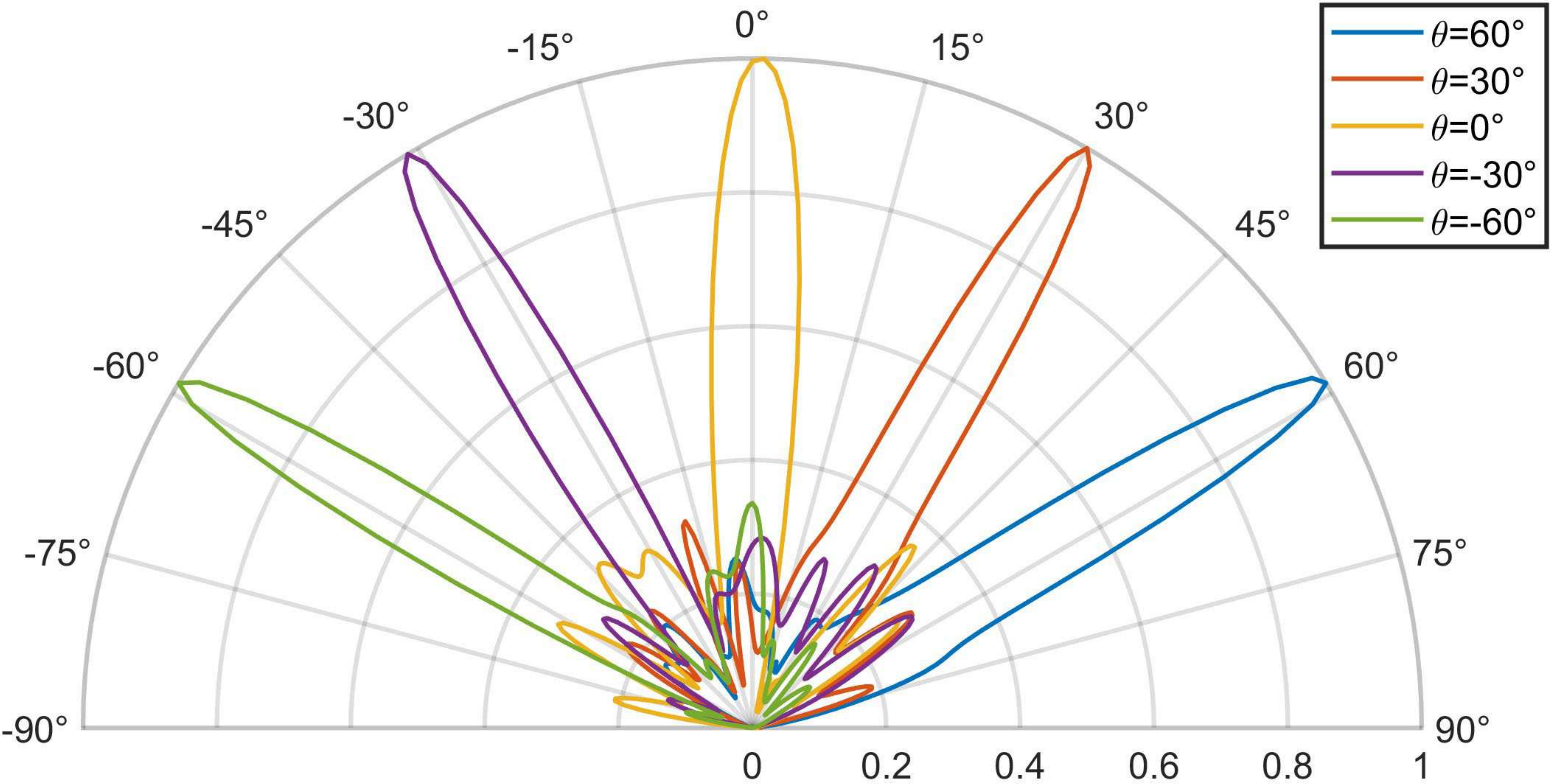}}
\hspace{0.01in}
\subfigure[Vertical plane]{
	\includegraphics[width=2.7in]{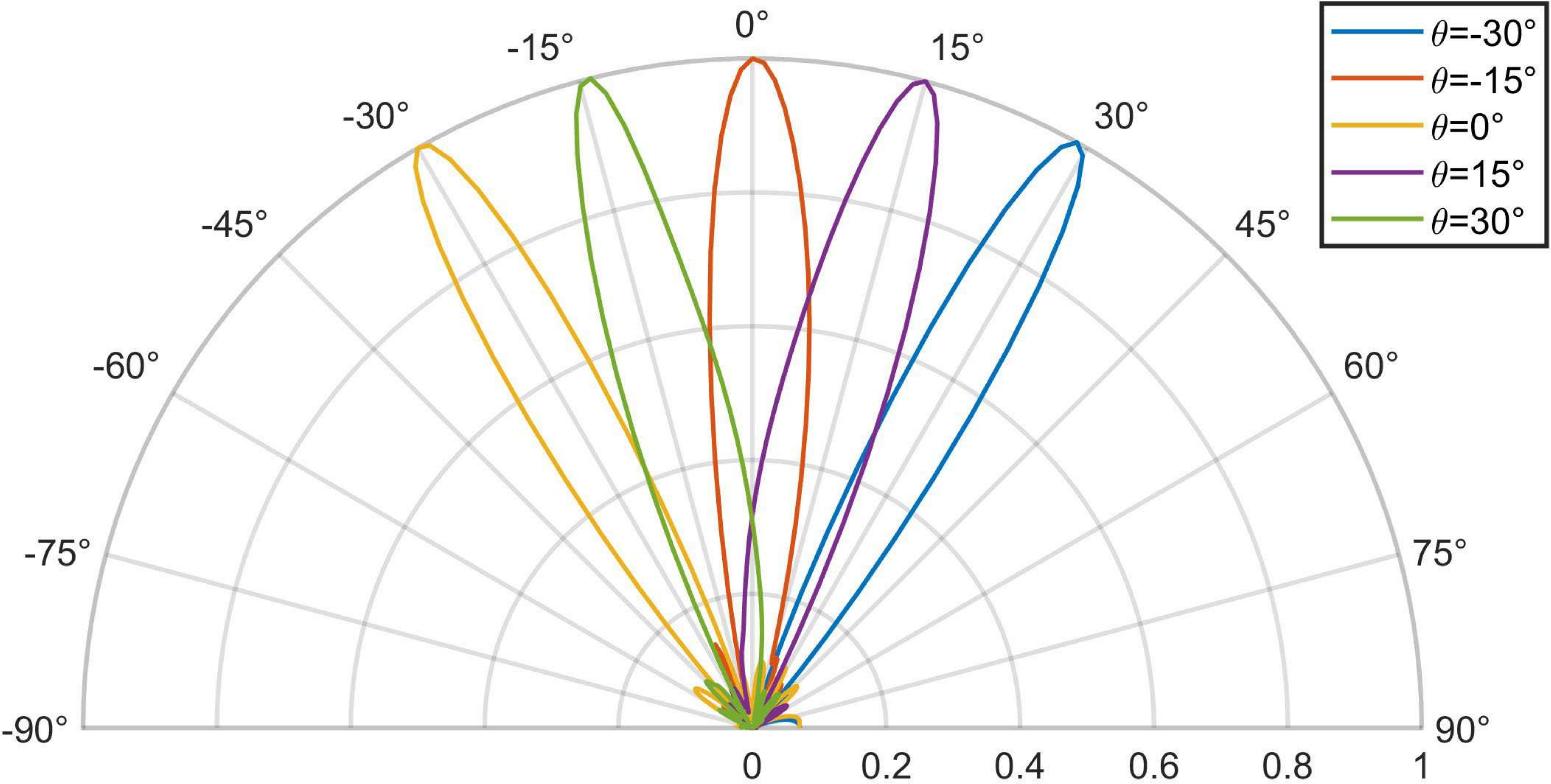}}
\caption{Normalized far-field beam patterns of the RHS in the horizontal plane and the vertical plane.}
\label{epa}
\end{figure}

The two-dimensional RHS have multiple feeds attached to the edge of the RHS. Note that the radiation amplitude of each RHS element is related to the distance between the RHS element and the feed according to (1). For a multi-feed two-dimensional RHS, the radiation amplitude of each RHS element can be calculated as a summation of the radiation amplitude distribution corresponding to each feed, i.e.,
\begin{equation}
m_n(\theta, \phi)=\sum_{k=1}^K\frac{\text{Re}[\Psi_{intf}(\mathbf{r}_n^k, \theta, \phi)]+1}{2K}.
\end{equation}

Fig.~3(a) and Fig.~3(b) show the normalized far-field beam pattern of the two-dimensional RHS with 64 elements in the horizontal plane with the desired direction scanned from $-60^{\circ}$ to $60^{\circ}$ and that in the vertical plane with the desired direction scanned from $-30^{\circ}$ to $30^{\circ}$, respectively\footnote{The sidelobe can be cancelled by superposing an auxiliary control pattern to reduce the holographic pattern corresponding to the sidelobe on the original holographic pattern~\cite{MSN-2016}.}. This validates the capability of the two-dimensional RHS to achieve three-dimensional holographic beamforming through the amplitude-controlled method. Moreover, the simulated gain of the RHS with 32 and 64 RHS elements is about 7 and 10~dBi, respectively. The simulated results show that the gain of the RHS will increase 3~dB when the number of RHS elements doubles, indicating that the RHS has the same gain-boosted capability as traditional antennas~\cite{DP-2005}. Hence, the following remark stands.

\textbf{Remark 2.} \emph{Capable of the same gain boosting as traditional antennas, the RHS has great potential to achieve a high directive gain through numerous RHS elements with low hardware cost. Hence, the RHS provides a practical way to enable holographic radio. }

Moreover, since the RHS adopts series feeding, the input signals are only required to be connected with the feeds rather than all RHS elements, leading to a simple wiring layout when implementing ultra-massive MIMO.

%


\section{Experimental Prototype of RHS-enabled Holographic Radio}\label{EP}
In this section, we introduce the implementation of the RHS prototype. The beam pattern measurement procedures of the prototype and the RHS-aided point-to-point communication platform are then presented.

\begin{figure}[t]
\centering
\includegraphics[width=3.2in]{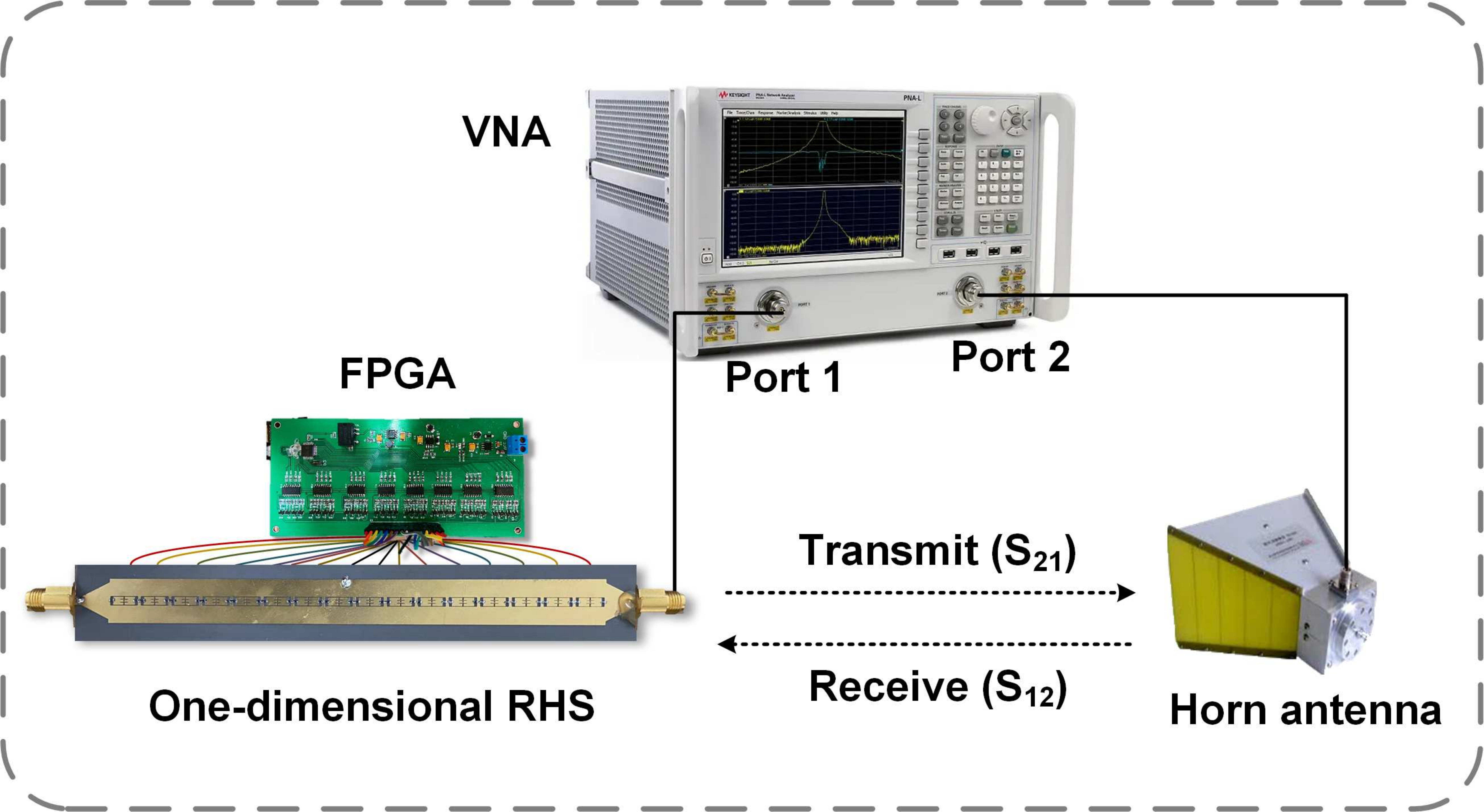}
\caption{Beam pattern measurement of the RHS.}
\vspace{-0.3cm}
\end{figure}

\begin{figure*}[t]
\centering
\includegraphics[width=6in]{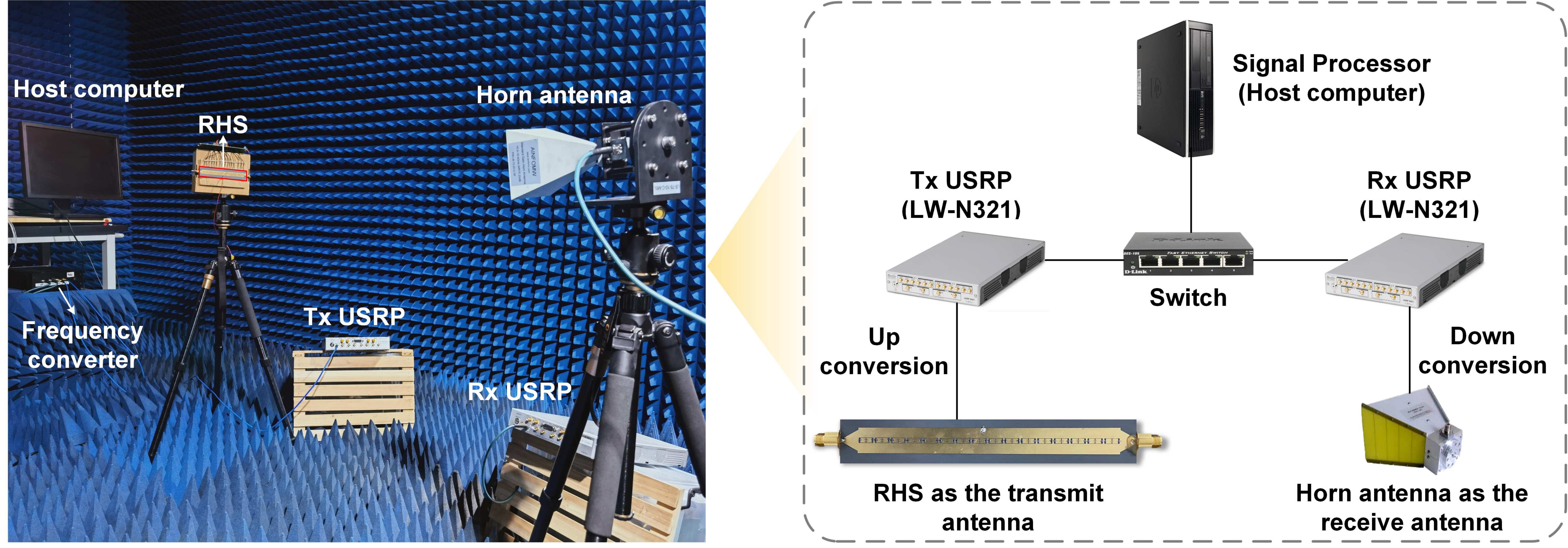}
\vspace{-0.1cm}
\caption{RHS-aided point-to-point communication platform.}
\vspace{-0.2cm}
\label{fig1}
\end{figure*}

\subsection{Implementation of the RHS Prototype}
As shown in Fig. 4, the designed one-dimensional RHS prototype consists of 16 RHS elements and the size of the RHS is $15.2\times 1.7\times 0.11$ $\text{cm}^3$. The radiation amplitude of each RHS element can be controlled based on a field-programmable gate array (FPGA). Specifically, the bias voltage applied to each PIN diode can be changed by controlling the FPGA, and thus, the ON/OFF state of the PIN diode can be controlled. In addition,  we utilize SubMiniature version A (SMA) connectors attached to the edge of the RHS to feed signals into the RHS and introduce a trapezoidal evolutionary structure to the edge of the RHS for impedance matching.


\subsection{Beam Pattern Measurement}


As shown in Fig. 4, a vector network analyzer (VNA) is utilized to measure the beam pattern of the RHS. Specifically, port 1 of the VNA is connected to the RHS, while port 2 of the VNA is connected to a standard horn antenna. The RHS is placed on an antenna rotating platform. To measure the transmit beam pattern of the RHS, the relative signal strength (i.e., $S_{21}$ parameter) received at the horn antenna corresponding to different angles is measured and normalized. To validate the transceiver reciprocity of the RHS, the receive beam pattern is obtained by measuring and normalizing the value of $|S_{12}|$ corresponding to different angles.


\vspace{-0.2cm}
\subsection{RHS-Aided Point-to-Point Communication Platform}
The hardware modules of the RHS-aided point-to-point communication platform are shown in Fig. 5 and their functions are introduced below.
\begin{itemize}
\item \emph{Transmitter:} The transmitter (Tx) is implemented by utilizing a universal software radio peripheral (USRP) LW-N321, which can realize RF modulation/demodulation and baseband signal processing. The output port of the USRP is connected to a frequency converter to up convert the signal frequency to 12~GHz. The frequency converter then sends the up-converted signal to the RHS.
\item \emph{Receiver:} The receiver (Rx) is also a USRP LW-N321, whose input port is connected to a frequency converter to down convert the received signal frequency. The frequency converter is connected to a standard horn antenna, which receives the signal emitted from the RHS.
\item \emph{Host computer:} The host computer can control the Tx and Rx via a software program. The received signal is also processed by the host computer. The FPGA can also transform the input from the host computer into different bias voltages applied to the PIN diodes of each RHS element.
\item \emph{Ethernet switch:} The ethernet switch connects the host computer, the Tx, and the Rx, and thus, the transmitted and received signals can be acquired.
\end{itemize}


\begin{figure}[t]
\vspace{-0.4cm}
\centering
\includegraphics[width=3in]{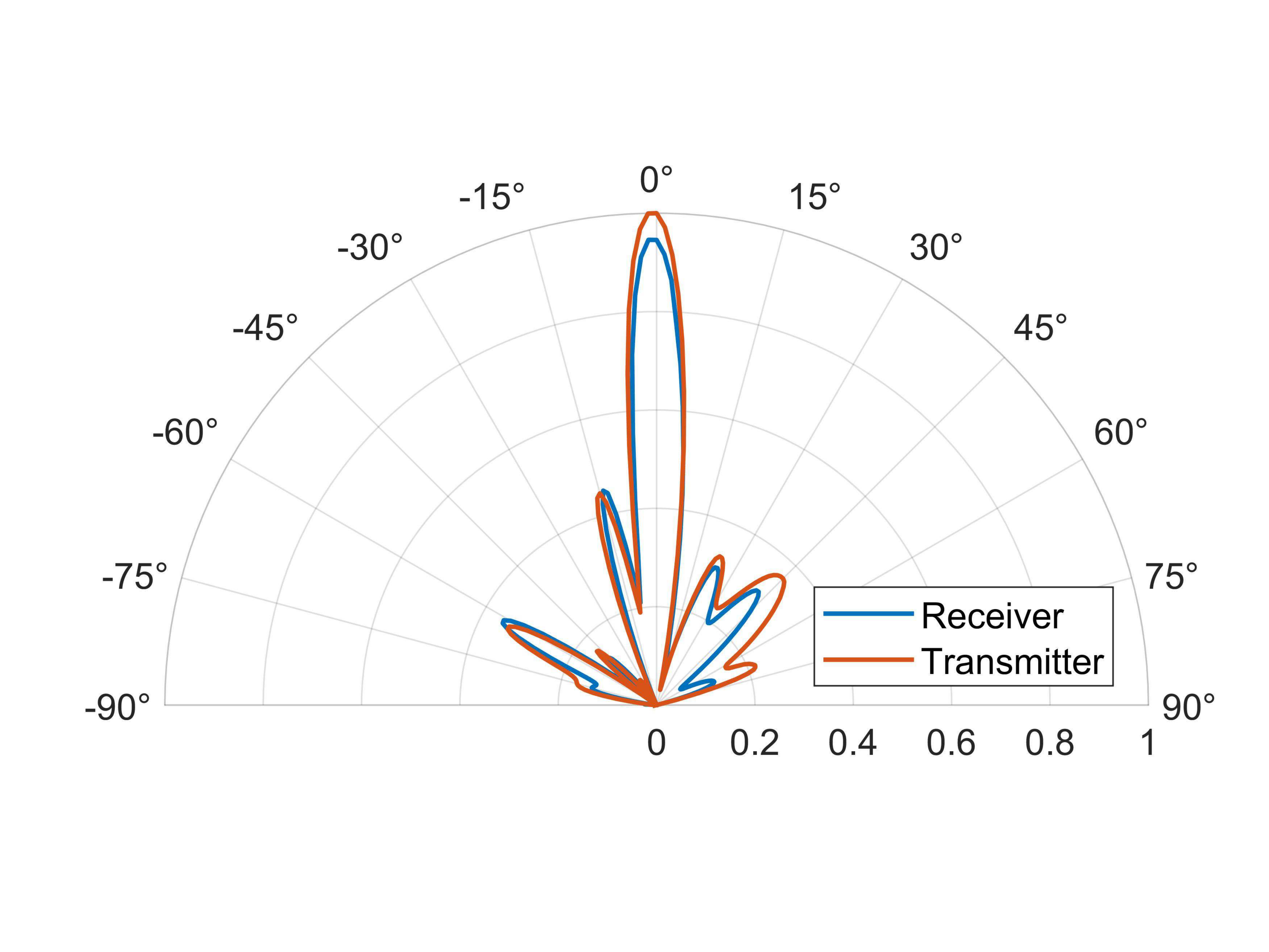}
\vspace{-1.2cm}
\caption{Normalized far-field beam pattern of the RHS.}
\vspace{-0.4cm}
\label{fig1}
\end{figure}

\section{Experimental Results and Discussion}\label{ER}
In this section, we present the experimental setup and experimental results obtained with the implemented RHS prototype.

\subsection{Experimental Setup}
To avoid environmental scatterings, the RHS prototype is deployed in a microwave anechoic chamber. To measure the far-field beam pattern of the RHS, the horn antenna acting as the receive/transmit antenna is placed 2 m away from the RHS\footnote{The Rayleigh distance of the RHS $d_{Ray}$ is $2D^2/\lambda$, where $D$ is the maximum dimension of the RHS, i.e., 15.2 cm, and $\lambda$ is the wavelength. Since the working frequency of the RHS is 12~GHz, $d_{Ray}$ is 1.8 m.}. In the deployed RHS-aided point-to-point communication platform, the
transmit power of the Tx USRP is 2~dBm, and the gain of the horn antenna is 20~dBi.

\vspace{-0.3cm}
\subsection{Experimental Results}

\begin{figure*}[t]
\centering
\includegraphics[width=6.4in]{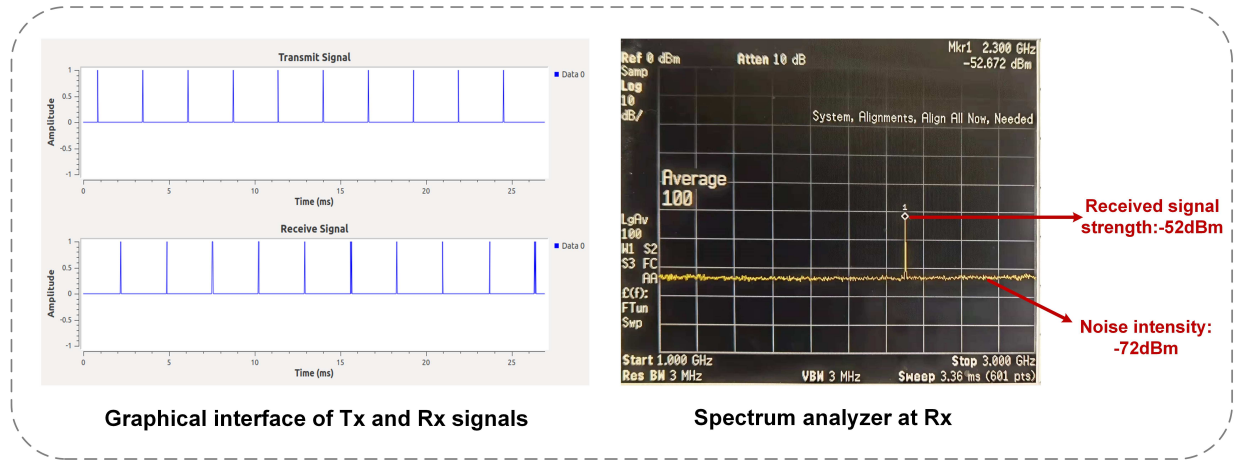}
\caption{Experimental results of RHS-aided point-to-point communications.}
\label{fig1}
\end{figure*}

Fig. 6 shows the normalized far-field beam pattern of the RHS. For convenience, we set the holographic pattern of the RHS generating the beam with the direction of $0^{\circ}$ according to~(1). Specifically, when the theoretical radiation amplitude of an RHS element is less than 0.5, the PIN diodes of the RHS element will be set in ON states and vice versa. It can be seen that the measured main lobe's direction of the beam pattern matches the object direction, indicating that holographic beamforming can be achieved based on~(1). We also observe that the transmit beam pattern and the receive beam pattern are almost the same, validating the transceiver reciprocity of the RHS. Moreover, the power consumption of the PIN diode in the implemented RHS is 0.01~W, which is far smaller than that of phase-shifting circuits. Hence, compared with traditional phased arrays, the RHS also provides a powerful solution to reduce power consumption in practice\footnote{More discussion and comparison about the performance of RHSs and traditional phased arrays can be found in~\cite{RBS-2021} and~\cite{RD-2021}, to which the interested readers are referred for further information.}.

Fig. 7 depicts the graphical interface of the Tx and Rx signals on the host computer when transmitting a real-time data stream. Specifically, the data stream generated by the vector source of the Tx USRP is a vector beginning with a 1 followed by a sequence of 0. We observe that the received signal demodulated by the Rx USRP is the same as the transmit signal. This validates that our RHS-aided point-to-point communication platform supports real-time data transmission. Moreover, from the received spectrum at the Rx, it can be seen that the signal-noise ratio (SNR) of the RHS-aided point-to-point communication is about 20 dB under the experimental environment.


\section{Future research directions of RHS-Enabled Holographic Radio}
In this section, we present future research directions for RHS-enabled holographic radio.

\vspace{-0.3cm}
\subsection{RHS-Aided Satellite Networks}

Low-Earth-orbit (LEO) satellite communication networks are being developed with the promise of providing high-capacity backhaul or data relay services for terrestrial networks. However, the high mobility of LEO satellites and the severe path loss put stringent requirements on antenna technologies in terms of accurate beam steering and high antenna gain. Traditional antennas integrated with UTs such as dish antennas and phased arrays either require heavy mechanics or costly phase shifters, making their implementation in practical systems prohibitive.

Since RHSs are ultra-thin and lightweight antennas and can achieve beamforming with low hardware cost and low power consumption, RHS-enabled holographic radio can be utilized in integrated terrestrial-satellite networks to overcome the shortfalls in traditional antennas. Since existing algorithms optimizing traditional complex-valued analog beamformer do not work well for real-valued holographic beamformer, a holographic beamforming scheme generating multiple directional beams toward the satellites needs to be developed for sum rate maximization. In addition, considering that the high mobility of LEO satellites leads to time-varying beam patterns, the RHS-aided multi-satellite communication protocol design is also worth exploring.

\vspace{0.2cm}
\subsection{Holographic Integrated Sensing and Communication}
ISAC, where the radar and communication systems share the common spectrum, is one of the most promising candidates to mitigate the spectrum congestion issue~\cite{HZ-2022}. However, the high power consumption and hardware cost of traditional phased arrays restrict the implementation of ultra-massive MIMO, leading to an insufficient ISAC performance.

Benefitting from the capability of achieving high directive gain with low hardware cost, RHS-enabled holographic radio can be applied in ISAC systems to enhance sensing and communication performance. In ISAC systems, the amplitude-controlled holographic beamforming design and the estimation method for target parameters are coupled with each other, which should be simultaneously optimized in the sensing scheme. Moreover, to effectively detect targets and serve communication users, a holographic beamforming optimization algorithm needs to be carefully designed while considering the trade-off between the sensing and communication functionalities.

\vspace{-0.6cm}
\subsection{RHS-Aided Wireless Simultaneous Localization and Mapping:}

Wireless SLAM, which estimates the position of the user and builds up the map of an unknown environment is a promising technique to empower location-based services. It utilizes antennas to estimate the time of arrival and the angle of arrival based on the amplitudes of received multipath components (MPCs). Hence, the accuracy of the SLAM system is determined by the directive gain of the antennas.

Considering that the RHS is composed of numerous RHS elements, leading to a superior beam-steering capability and high directive gain, the RHS can be utilized in wireless SLAM systems. By adjusting the amplitude of each RHS element properly, the amplitudes of MPCs can be enhanced, such that the accuracy of the wireless SLAM system can be improved.

\vspace{0.2cm}
\section{Conclusions}
In this article, we have investigated RHS-enabled holographic radio for 6G communications. The working principle of RHSs including holographic beamforming has been introduced. The hardware design of the RHS element and full-wave analyses of both single-feed one-dimensional RHS and multi-feed two-dimensional RHS have then been presented. Notably, we have proposed a prototype of RHS-enabled holographic radio where an RHS-aided point-to-point communication platform supporting real-time data transmission has been built. It is proved that due to diode-based controllers and series feeding, the RHS has great potential to achieve high directive gain with low hardware cost, low power consumption, and a simple wiring layout, thereby providing a practical way toward ultra-massive MIMO. Future research directions for RHS-enabled holographic radio including satellite communications, ISAC, and SLAM have also been discussed.


\vspace{-0.2cm}
\begin{thebibliography}{30}

\bibitem{EOF-2014}
E. G. Larsson, O. Edfors, F. Tufvesson, and T. L. Marzetta, ``Massive MIMO for next generation wireless systems," \emph{IEEE Commun. Mag.}, vol. 52, no. 2, pp. 186-195, Feb. 2014.

\bibitem{ZZF-2021}
Z. Wan, Z. Gao, F. Gao, M. Di Renzo, and M.-S. Alouini, ``Terahertz massive mimo with holographic reconfigurable intelligent surfaces,"
\emph{IEEE Trans. Commun.}, vol. 69, no. 7, pp. 4732-4750, Mar. 2021.

\bibitem{ATL-2020}
C. Huang \emph{et al.}, ``Holographic MIMO surfaces for 6G wireless networks: Opportunities, challenges, and trends," \emph{IEEE Wireless Commun.}, vol. 27, no. 5, pp. 118-125, Oct.~2020.

\bibitem{BFY-2020}
R. B. Hwang, ``Binary meta-hologram for a reconfigurable holographic metamaterial antenna," \emph{Sci. Rep.}, vol. 10, no. 1, pp. 1-10, May. 2020.


\bibitem{BHL-2020}
C. Huang, A. Zappone, G. C. Alexandropoulos, M. Debbah, and C. Yuen, ``Reconfigurable intelligent surfaces for energy efficiency in wireless communication," \emph{IEEE Trans. Wireless Commun.}, vol. 18, no. 8, pp. 4157-4170, Aug. 2019.

\bibitem{MSJ-2014}
B. Che, F. Meng, Y. Lyu, Y. Zhao and Q. Wu, ``Reconfigurable holographic antenna with low sidelobe level based on liquid crystals," \emph{J. Phys. D: Appl. Phys.}, vol. 53, no. 31, May 2020.



\bibitem{T-2016}
T. Sleasman \emph{et al.}, ``Waveguide-fed tunable metamaterial element for dynamic apertures," \emph{IEEE Antennas Wireless Propag. Lett.}, vol. 15, pp.~606-609, July 2016.

\bibitem{RBS-2021}
R. Deng, B. Di, H. Zhang, Y. Tan, and L. Song, ``Reconfigurable holographic surface enabled multi-user wireless communications: Amplitude-controlled holographic beamforming," \emph{IEEE Trans. Wireless Commun.}, vol. 21, no. 8, pp. 6003-6017, Aug. 2022.

\bibitem{NOY-2019}
N. Shlezinger, O. Dicker, Y. C. Eldar, I. Yoo, M. F. Imani, and D. R. Smith, ``Dynamic metasurface antennas for uplink massive MIMO systems," \emph{IEEE Trans. Commun.}, vol. 67, no. 10, pp. 6829-6843, Oct. 2019.


\bibitem{DO-2016}
D. R. Smith, O. Yurduseven, L. P. Mancera, P. Bowen, and N. B. Kundtz, ``Analysis of a waveguide-fed metasurface antenna," \emph{Physical Review Applied}, vol. 8, no. 5, pp. 054048, Nov. 2017.



\bibitem{MSN-2016}
M. Johnson, S. Brunton, N. Kundtz, and N. Kutz, ``Extremum-seeking control of the beam pattern of a reconfigurable holographic metamaterial antenna," \emph{J. Opt. Soc. Amer. A}, vol. 33, no. 1, pp. 59-68, Jan. 2016.


\bibitem{HY-2016}
H. Yang \emph{et al.}, ``A 1-Bit $10 \times 10$ reconfigurable reflectarray antenna: Design, optimization, and experiment," \emph{IEEE Trans. Antennas Propag.}, vol. 64, no. 6, pp. 2246-2254, June 2016.

\bibitem{RD-2021}
R. Deng, B. Di, H. Zhang, H. V. Poor, and L. Song, ``Holographic MIMO for LEO satellite communications aided by reconfigurable holographic surfaces," \emph{IEEE J. Sel. Areas Commun.}, early access.

\bibitem{DP-2005}
D. Tse and P. Viswanath, Fundamentals of Wireless Communications. Cambridge, U.K.: Cambridge Univ. Press, 2005.

\bibitem{HZ-2022}
H. Zhang \emph{et al.}, ``Holographic integrated sensing and communication," \emph{IEEE J. Sel. Areas Commun.}, early access.


\end{thebibliography}
\end{document}